\documentclass{emulateapj}
\usepackage{latexsym}
\usepackage{mathrsfs}
\usepackage{amsfonts}
\usepackage{amssymb}
\usepackage{amsmath}
\usepackage{natbib}
\usepackage{subfig}

\newcommand{\be}{\begin{equation}}
\newcommand{\ee}{\end{equation}}
\newcommand{\bes}{\begin{equation*}}
\newcommand{\ees}{\end{equation*}}
\newcommand{\bsy}{\boldsymbol}

\shorttitle{Coronal Heating}
\shortauthors{Rappazzo et al.}

\begin{document}
\title{Coronal Heating, Weak MHD Turbulence and Scaling Laws}
\author{A. F. Rappazzo \altaffilmark{1,2},
M. Velli \altaffilmark{2,3},
G.~Einaudi \altaffilmark{1} and
R.~B. Dahlburg \altaffilmark{4} 
}

\altaffiltext{1}{Dipartimento di Fisica ``E.~Fermi'', Universit\`a di Pisa,
                        56127 Pisa, Italy; rappazzo@jpl.nasa.gov}
\altaffiltext{2}{Jet Propulsion Laboratory, Pasadena, CA,
                        91109}
\altaffiltext{3}{Dipartimento di Astronomia e Scienza dello Spazio, 
                        Universit\`a di Firenze, 50125 Firenze, Italy}
\altaffiltext{4}{LCP\&FD, Naval Research Laboratory, 
                        Washington, DC 20375}
       
\begin{abstract}
Long-time high-resolution simulations of the dynamics of a coronal loop in 
cartesian geometry are carried out, within the framework of reduced 
magnetohydrodynamics (RMHD), to understand coronal heating driven by motion of 
field lines anchored in the photosphere. 
We unambiguously identify  MHD anisotropic turbulence as the physical mechanism
responsible for the transport of energy from the large scales, where energy is injected
by photospheric motions,  to the small scales, where it is dissipated.
As the loop parameters vary  different regimes of turbulence develop: 
strong turbulence is found for weak axial magnetic fields and long loops,
leading to Kolmogorov-like spectra in the perpendicular direction, while 
weaker and weaker regimes (steeper spectral slopes of total energy) are 
found for strong axial magnetic fields and short loops. 
As a consequence we predict that the  scaling of  the heating rate with 
axial magnetic field intensity $B_0$,
which depends on the spectral index of total energy for given loop 
parameters, must vary from $B_0^{3/2}$ for weak fields to 
$B_0^{2}$ for 
strong fields at a given aspect ratio. The predicted heating rate is within the 
lower range of observed active region and quiet Sun coronal energy losses.
\end{abstract}

\section{INTRODUCTION}

In this letter we solve, within the framework of RMHD in cartesian 
geometry, the Parker field-line tangling (coronal heating) 
problem \citep{park72,park88}.
We do this via long simulations at high resolutions, introducing 
hyper-resistivity models to attain extremely large Reynolds 
numbers. We show how small scales form and how the coronal 
heating rate depends on the loop and photospheric 
driving parameters, and derive simple formulae which may be used in 
the coronal heating context for other stars. 

Over the years a number of numerical experiments have been carried out to 
investigate coronal heating, with particular emphasis on exploring  how photospheric 
field line tangling leads to current sheet formation.

\citet{mik89} and \citet{hen96} first carried out simulations 
of a loop driven by photospheric motions using a cartesian approximation (a 
straightened out loop bounded at each end by the photosphere) imposing a 
time-dependent alternate direction flow pattern at the boundaries.  
A complex coronal magnetic field results  from the photospheric field line random walk, and though the  field  does not, strictly speaking, evolve through a sequence of static force-free 
equilibrium states (the original Parker hypothesis), magnetic energy nonetheless
tends to dominate kinetic energy in the system. 
In this limit the field is structured by current sheets elongated along the axial 
direction,  separating quasi-2D flux tubes which constantly move around and interact.  
\citet{long94}  focused on the current sheet formation process within 
the RMHD approximation, also used in the simulations by
\citet{dmi99}. The results from these studies agreed qualitatively among 
themselves, in that all simulations display the development of field aligned 
current sheets. However, estimates of the dissipated power and its scaling 
characteristics differed largely, depending on the way in which
extrapolations from low to large values of the plasma conductivity 
of the properties such as  inertial range power law 
indices were carried out.
2D numerical simulations of incompressible MHD with magnetic forcing 
\citep{ein96,georg98,dmi98,ein99} showed that turbulent current sheets 
dissipation is distributed intermittently, and that the
statistics of dissipation events, in terms of total energy, peak energy and event duration
displays power laws not unlike
the distribution of observed emission events in optical, ultraviolet and x-ray 
wavelengths of the quiet solar corona.

More recently full 3D sections of the solar corona with a realistic geometry have been 
simulated by \citet{gud05}. While this approach has 
advantages when investigating the coronal loop dynamics within its neighboring 
coronal region, modeling numerically a larger part of the solar corona drastically reduces 
the number of points occupied by the coronal loops. Thus, these  simulations 
have not been able to shed further light on the \emph{physical mechanism} 
responsible for the coronal heating.

In \S~\ref{model} we introduce the coronal loop 
model and the simulations we have carried out; 
in \S~\ref{results} we describe
our numerical results, and in \S~\ref{disc} we give simple scaling arguments 
to understand the magnetic energy spectral slopes. This will lead to a 
quantitative asymptotic estimate of the coronal loop heating rate, and of its 
scaling with the axial magnetic field, photospheric velocity amplitude 
and coronal loop length. 
\begin{figure}
     \includegraphics[width=0.47\textwidth]{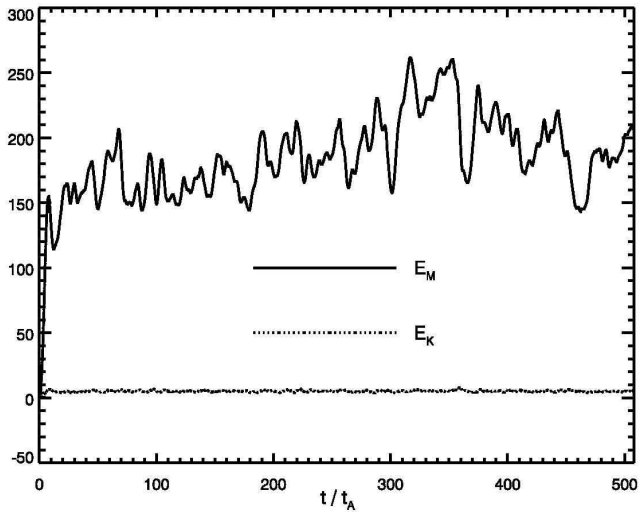}
      \caption{High-resolution simulation with 
                $v_{\mathcal A}/u_{ph} = 200$, 512x512x200 
                grid points and $\mathcal{R}_1=800$.
                Magnetic ($E_M$) and kinetic ($E_K$) energies as a function 
                of time ($\tau_{\mathcal A}=L/v_{\mathcal A}$ is the axial Alfv\'enic 
                crossing time).  
     \label{en}}
\end{figure}

\section{THE MODEL} \label{model}

A coronal loop is a closed magnetic structure threaded by a  strong axial 
field, with the footpoints rooted in the photosphere.
This makes it a strongly anisotropic system, as measured by 
the relative magnitude of the Alfv\'en velocity 
$v_{\mathcal A} \sim 1000\ \textrm{km}\, \textrm{s}^{-1}$ compared 
to the typical photospheric velocity
$u_{ph} \sim 1\ \textrm{km}\, \textrm{s}^{-1}$. 
This means that the relative amplitude of the
Alfv\'en waves that are launched into the corona is very
small. The loop dynamics 
may be studied in 
a simplified geometry, neglecting any curvature effect,  as a 
``straightened out'' cartesian box,  
with an orthogonal square cross section of 
size $\ell_{\perp}$, and an 
axial length $L$ embedded in an axial homogeneous uniform magnetic field 
$\boldsymbol{B}_0 = B_0\ \boldsymbol{e}_z$. This system may be described 
by the reduced MHD (RMHD) equations 
\citep{kp74,stra76}: introducing the velocity and magnetic field 
potentials $\varphi$ and $\psi$,
$\boldsymbol u_\perp = \boldsymbol{\nabla} \times 
\left( \varphi\, \boldsymbol{e}_z \right)$,
$\boldsymbol b_\perp = \boldsymbol{\nabla} \times 
\left( \psi\, \boldsymbol{e}_z \right)$, and vorticity and current, 
$\omega =- \boldsymbol{\nabla}^2_\perp \varphi$,
$j = - \boldsymbol{\nabla}^2_\perp \psi$ the non-dimensioned RMHD system is  
given by
\begin{eqnarray}
& &\frac{\partial \psi}{\partial t} = v_{\mathcal A}\, 
\frac{\partial \varphi}{\partial z} + 
\left[ \varphi, \psi \right] + \frac{\left(-1\right)^{n+1}}{\mathcal{R}_n} \boldsymbol{\nabla}^{2n}_\perp 
\psi,  \label{pot1} \\
& &\frac{\partial \omega}{\partial t} = v_{\mathcal A}\, 
\frac{\partial j}{\partial z} + 
\left[ j , \psi \right] - \left[ \omega , \varphi \right] 
+ \frac{\left(-1\right)^{n+1}}{\mathcal{R}_n} \boldsymbol{\nabla}^{2n}_\perp \omega. \label{pot2} 
\end{eqnarray}
As characteristic quantities we use the perpendicular length of the computational
box $\ell_{\perp}$, the typical photospheric velocity $u_{ph}$, and the related
crossing time $t_{\perp} = \ell_{\perp} / u_{ph}$. The equations have been 
rendered dimensionless using velocity units for the magnetic field (the density in the 
loops $\rho$ is taken to be constant) and normalizing by $u_{ph}$.
Then the non-dimensioned Alfv\'en speed $v_{\mathcal A}$ in eqs.~(\ref{pot1})-(\ref{pot2})
is given by the ratio $v_{\mathcal A}/u_{ph}$ between the dimensional velocities.
The Poisson bracket of two functions $g$ and $h$ is defined as
$\left[ g, h \right] = \partial_x g\, \partial_y h - \partial_y g\, \partial_x h$,
where $x,y$ are transverse coordinates across the loop while $z$ is the axial 
coordinate along the loop. A simplified diffusion model is assumed and
$\mathcal{R}_n$ is the Reynolds number, with $n$ the hyperdiffusion index
(\emph{dissipativity}): for $n=1$ ordinary diffusion is recovered.

The computational box spans $ 0 \le x, y \le 1 $ and $ 0 \le z \le L$, 
with $L=10$, corresponding to an aspect ratio equal to $10$.
As boundary conditions at the photospheric surfaces 
($z=0,\ L$) we impose a velocity pattern intended to
mimic photospheric motions, made up of two independent large spatial 
scale projected convection cell flow patterns.  The wave number values 
$k$ excited are all those in the range $3 \le k \le 4$, and the 
average injection wavenumber is $k_{in} \sim 3.4$.
\begin{figure}
     \includegraphics[width=0.47\textwidth]{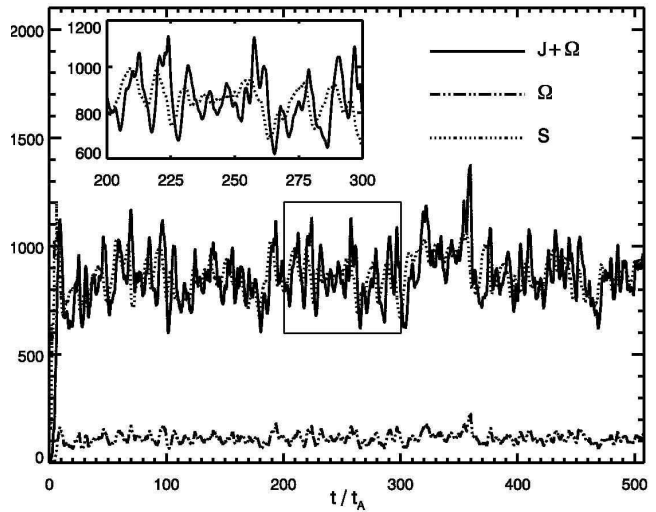}
      \caption{Same simulation of Figure~\ref{en}.
                 The integrated Poynting flux $S$ dynamically balances the 
                 Ohmic ($J$) and viscous ($\Omega$) dissipation.
                 Inset shows a magnification of total dissipation and
                 $S$ for $200 \le t/\tau_{\mathcal{A}} \le 300$.
     \label{diss}}
\end{figure}

\section{RESULTS} \label{results}

Plots of the rms magnetic and kinetic energies as a function of time, 
together with 
the dissipation due to currents, vorticity, as well as the integrated 
Poynting flux, 
are shown in Figures~\ref{en} and \ref{diss}. As a result of the photospheric forcing, 
energy in the magnetic 
field first grows with time, until it dominates over the kinetic energy 
by a large 
factor, before oscillating, chaotically, around a stationary state.  Fluctuating 
magnetic energy $E_M$ is $\sim 35$ times bigger than kinetic energy $E_K$.
\begin{figure} \label{fig:hypsp}
\includegraphics[width=0.45\textwidth]{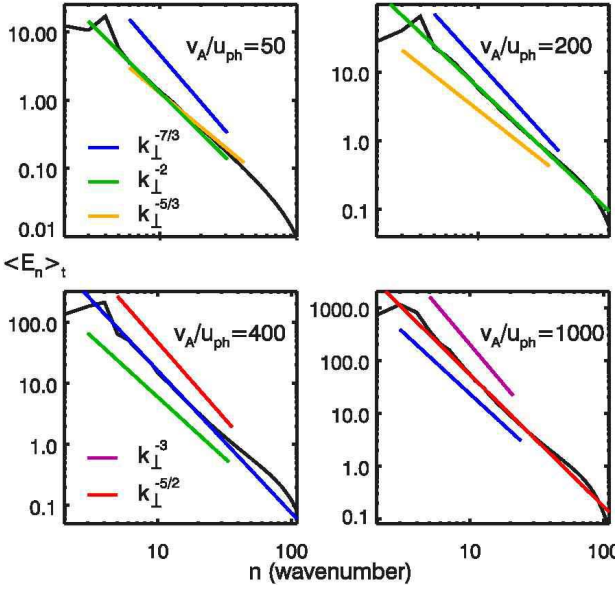}
      \caption{Time-averaged total energy spectra
      for simulations with
      $v_{\mathcal A}/u_{ph} = 50,\ 200,\ 400,\ 1000$. Hyperdiffusion ($n=4$) 
      has been used  with 
      $\mathcal{R}_4 = 3 \cdot 10^{20}$, $10^{20}$, $10^{19}$,
      $10^{19}$ respectively, and a grid with
      $512\times512\times200$ points.}
\end{figure}

\begin{figure}[p]
      \centering
      \subfloat{
               \includegraphics[height=0.56\linewidth]{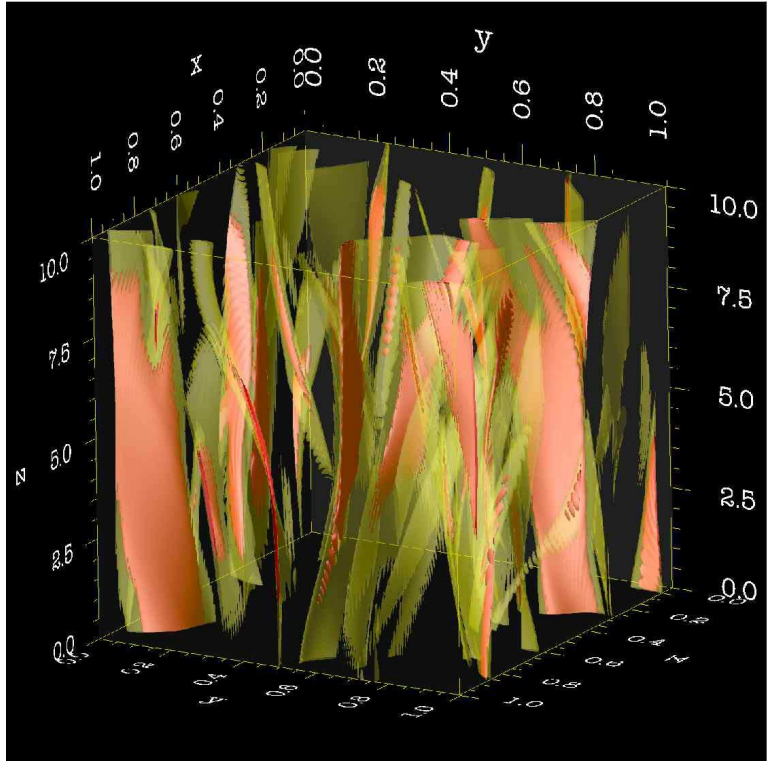}}\\[20pt]
      \subfloat{
               \includegraphics[height=0.56\linewidth]{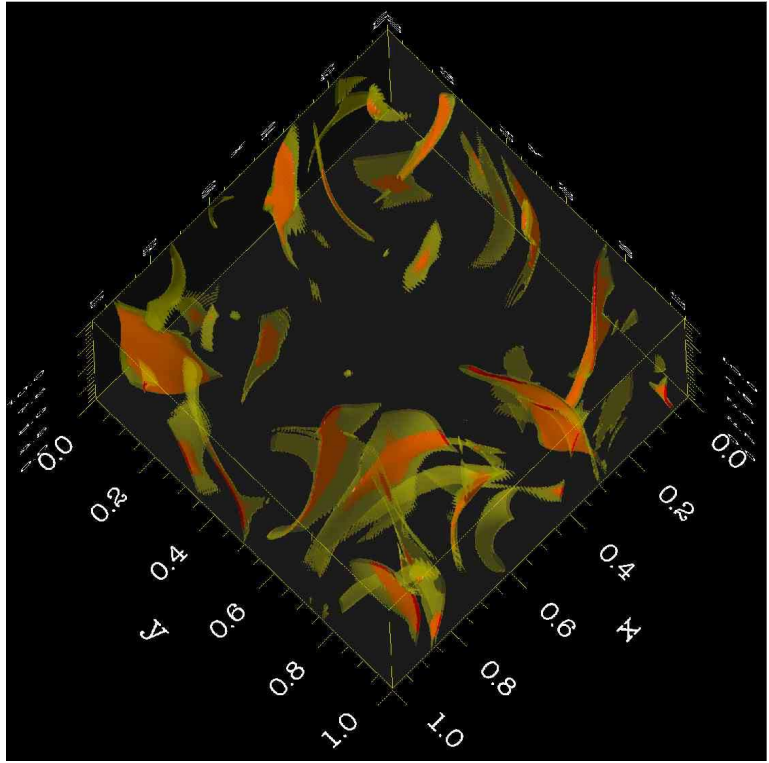}}
      \caption{\emph{Top}: side view of 
               two isosurfaces of the squared current at a selected time
               for a numerical simulation with $v_{\mathcal A}/u_{ph} = 200$,
                512x512x200 grid points and a Reynolds
               number $\mathcal R_1 =800$.
               The isosurface at the value $j^2 = 2.8 \cdot 10^5$ is represented in 
               partially
               transparent yellow,  while  red displays
               the isosurface with  $j^2 = 8 \cdot 10^5$, well below the value of the
               maximum of the squared current that at this time is 
               $j^2 = 8.4 \cdot 10^6$.
               N.B.: The red isosurface is always nested inside the yellow one, and
               appears pink in the figure. The computational box has been rescaled
               for an improved viewing, but the aspect ratio of the box is $10$,
               i.e.\ the axial length of the box is ten times bigger than
               its orthogonal length.
               \emph{Bottom}: top view  of the same two isosurfaces
                using the same color display.
                The isosurfaces are extended along the axial direction, and the
                corresponding filling factor is small.}
                \label{isofig}
\end{figure}

The same generic features are seen in the rms current and vorticity dissipation, 
where however the time dependence of the signal is more strongly oscillating. 
The ohmic dissipation rate $J$ is $\sim 6.5$ times viscous dissipation $\Omega$. 
The Poynting flux, on average, follows the current dissipation (there is no 
accumulation of energy in the box), however a detailed  examination shows that  
the dissipation time-series tends to lag the Poynting flux, with notable 
de-correlations around significant dissipation peaks. 
The spatial configuration of the currents which corresponds to a snapshot at a 
given time is displayed in Figure~\ref{isofig}. The currents
collapse into warped, torn sheets 
which extend almost completely along the loop. The current peaks are embedded 
within the 2D sheet-like structures,  corresponding to an anisotropic structure 
for the turbulence, in agreement with previous results. 

A dimensional analysis of eqs.~(\ref{pot1})-(\ref{pot2}) shows that
the only free nondimensional quantity is $f = \ell_{\perp} v_{\mathcal A} /L u_{ph}$. 
We fix
$L / \ell_{\perp} = 10$ and 
vary the ratio  of the Alfv\'en speed to photospheric 
convection speed $v_{\mathcal A} / u_{ph}$.
Both runs with standard second order dissipation ($n=1$) as well as
hyperdiffusion ($n=4$) have been carried out to obtain extended inertial ranges in the 
resulting spectra.

The power spectrum of total energy in the 
simulation box, once a statistically stationary state has been achieved, 
depends strongly on the ratio $v_{\mathcal A}/u_{ph}$.
This was first found in simulations by \citet{dmi03},
devoted to understanding how anisotropic regimes of MHD turbulence depend on boundary driving strength, with whom our numerical work is in broad agreement.

The total energy spectrum, for values of 
$v_{\mathcal A}/u_{ph} = 50,\ 200,\ 400,\ 1000$ is shown in 
Figure~\ref{fig:hypsp}, together with fits to the inertial range power law. As 
$v_{\mathcal A}/u_{ph}$ increases, 
the spectrum steepens visibly (note that the hump at the high wave-vector values 
for the runs with large $v_{\mathcal A}/u_{ph}$ is a feature, the bottleneck effect, which is well 
known and documented in spectral simulations of turbulence with the hyperdiffusion
used here, e.g.\ \citet{falk94}), with the slopes ranging from $-2$ to almost $-3$. At the same time
while total energy increases, the ratio of the mean magnetic field over the
axial Alfv\'en velocity decreases, in good accordance with the theory.
This steepening, which may be 
interpreted both as the effect of inertial line-tying of the 
coronal magnetic field and the progressive weakening of non-linear interactions 
as the magnetic field is increased, has a strong and direct bearing on the coronal heating 
scaling laws.

\section{DISCUSSION} \label{disc}

A characteristic of anisotropic MHD turbulence is that
the cascade takes place mainly in the plane orthogonal
to the DC magnetic guide field \citep{sheb83}.
Consider then the anisotropic version of the 
Iroshnikov-Kraichnan (IK) theory \citep{gold94,gold97}. 
Dimensionally the energy cascade rate 
may be written as
$\rho\, {\delta z_{\lambda}}^2 / T_{\lambda}$,
where $\delta z_{\lambda}$ is the rms value of the Els\"asser fields
$\bsy{z}^{\pm} = \bsy{u}_{\perp} \pm \bsy{b}_{\perp}$ at the 
perpendicular scale $\lambda$, where because the system is 
magnetically dominated $\delta z^+_{\lambda} \sim \delta z^-_{\lambda}$. 
$\rho$ is the average density and 
$T_{\lambda}$ is the energy transfer time
at the scale $\lambda$, which is greater than the eddy turnover time 
$\tau_{\lambda} \sim \lambda / \delta z_{\lambda}$ because of the
Alfv\'en effect \citep{iro64,kra65}. 

In the classical IK case,  
$T_{\lambda}\sim \tau_{\mathcal A} \bigl(\tau_{\lambda}/\tau_{\mathcal A}\bigr)^2$. 
This corresponds to the fact that wave-packets interact over an 
Alfv\'en crossing time (with $\tau_{\lambda} > \tau_{\mathcal A}$), 
and the collisions follow a standard random walk in energy exchange. 
In terms of the number of collisions $N_{\lambda}$ that a wave packet must 
suffer for the perturbation to build up to order unity, for IK
$N_{\lambda} \sim (\tau_{\lambda} / \tau_{\mathcal A})^2$. 
 
More generally, however, as the Alfv\'en speed is increased the interaction
time becomes smaller, so that  turbulence becomes weaker and the number 
of collisions required for efficient energy transfer scales as
\be \label{eq:atnc}
N_{\lambda} = \left( \frac{\tau_{\lambda}}{\tau_{\mathcal A}} \right)^{\alpha}
\qquad \mathrm{with} \qquad \alpha > 2,
\ee
where $\alpha$ is the scaling index
(note that $\alpha =1$ corresponds to standard hydrodynamic turbulence),
so that 
\be \label{eq:btnc}
T_{\lambda} \sim N_{\lambda}\, \tau_{\mathcal A} \sim
\left( \frac{v_{\mathcal A}}{L} \right)^{\alpha -1} 
\left( \frac{\lambda}{\delta z_{\lambda}} \right)^{\alpha}.
\ee
Integrating over the whole volume, the energy transfer rate becomes
\be \label{eq:sbe1}
\epsilon \sim \ell_{\perp}^2 L\cdot \rho\, \frac{\delta z_{\lambda}^2}{T_{\lambda}} 
\sim \ell_{\perp}^2 L\cdot \rho\, \left( \frac{L}{v_{\mathcal A}} \right)^{\alpha - 1} \, 
\frac{\delta z_{\lambda}^{\alpha + 2}}{\lambda^{\alpha}}.
\ee
Considering the injection scale $\lambda \sim \ell_{\perp}$,
eq.~(\ref{eq:sbe1}) becomes
\be \label{eq:sbe2}
\epsilon 
\sim \ell_{\perp}^2 L\cdot \rho\, \frac{\delta z_{\ell_{\perp}}^2}{T_{\ell_{\perp}}} 
\sim \frac{\rho \ell_{\perp}^2 L^{\alpha}}{\ell_{\perp}^{\alpha} \, v_{\mathcal A}^{\alpha - 1}} \, 
\delta z_{\ell_{\perp}}^{\alpha + 2}.
\ee
On the other hand the energy injection rate is given by the 
Poynting flux integrated across the photospheric boundaries: 
$\epsilon_{in} = \rho\, v_{\mathcal A} 
\int \! \mathrm{d} a\, \bsy{u}_{ph} \cdot \bsy{b}_{\perp}$.
Considering that  this integral is dominated by energy at the
large scales, due to the characteristics of the forcing function, we can approximate it with 
\be \label{eq:pf}
\epsilon_{in} \sim \rho\, \ell_{\perp}^2 v_{\mathcal A} u_{ph} \delta z_{\ell_{\perp}},
\ee
where the large scale component of the magnetic
field can be replaced with $\delta z_{\ell_{\perp}}$ because the system is magnetically dominated.

The last two equations show that the system is self-organized because 
both $\epsilon$ and $\epsilon_{in}$ depend on $\delta z_{\ell_{\perp}}$, 
the rms values of the fields
$\boldsymbol{z}^{\pm}$ at the scale $\ell_{\perp}$:
the internal dynamics depends
on the injection of energy and the injection of energy itself depends 
on the internal dynamics via the boundary forcing. 
Another aspect of self-organization results from our simulations: 
the perpendicular magnetic field develops few spatial structures
along the axial direction $z$, and in the nonlinear stage 
its topology substantially departs from the mapping of
the boundary velocity pattern which characterizes its evolution
during the linear stage. These and other features will be 
discussed more in depth in \citet{rapprep}.

In a stationary cascade the injection rate (\ref{eq:pf}) is equal to 
the transport rate (\ref{eq:sbe2}). Equating the two yields for 
the amplitude at the scale $\ell_{\perp}$:
\be \label{eq:amp}
\frac{\delta z_{\ell_{\perp}}^{\ast}}{u_{ph}} 
\sim \left( \frac{\ell_{\perp} v_{\mathcal A}}{L u_{ph}} \right)^{\frac{\alpha}{\alpha + 1}}
\ee
Substituting this value in (\ref{eq:sbe2}) or (\ref{eq:pf}) we obtain
for the energy flux
\be \label{eq:chs}
\epsilon^{\ast} 
\sim \ell_{\perp}^2 \, \rho \, v_{\mathcal A} u_{ph}^2 
\left( \frac{\ell_{\perp} v_{\mathcal A}}{L u_{ph}} \right)^{\frac{\alpha}{\alpha+1}},
\ee
where $v_{\mathcal A} = B_0 / \sqrt{4\pi\rho}$.
This is also the dissipation rate, and hence the \emph{coronal heating scaling}.
The crucial parameter here is $f = \ell_{\perp} v_{\mathcal A}/ L u_{ph}$ 
because the scaling index $\alpha$~(\ref{eq:atnc}), upon which the strength of 
the stationary turbulent 
regime depends, must be a function of $f$ itself. The relative amplitude of 
the turbulence $\delta z_{\ell_{\perp}}^{\ast} / v_{\mathcal A}$, is a function of $f$, 
and as
$f$ increases the effect of \emph{line-tying} becomes stronger, decreasing the 
strength of turbulent interactions (wave-packet collision efficiency becomes 
sub-diffusive) so that $\alpha$ increases above $2$. 
The ratio $\delta z_{\ell_{\perp}}^{\ast} / v_{\mathcal A}$
can also be interpreted as the rms value of the
Parker angle $\Theta_P$,
and is given by
\be \label{pan}
< \Theta_P > 
\sim \frac{\delta z_{\ell_{\perp}}^{\ast}}{v_{\mathcal A}} 
\sim \left( \frac{\ell_{\perp}}{L} \right)^{\frac{\alpha}{\alpha + 1}}
\left( \frac{u_{ph}}{v_{\mathcal A}} \right)^{\frac{1}{\alpha + 1}}.
\ee
This is actually an estimate of the average inclination of the magnetic field
lines,  while the rms value of the shear  angle between 
neighboring field lines is at least twice that given by eq.~(\ref{pan}),
not considering that close to a current sheet an enhancement of 
the orthogonal magnetic field is observed (which leads to a 
higher value for the angle).

Numerical simulations determine the remaining unknown nondimensional 
dependence of the scaling index $\alpha$ on $f$. The power law slopes of the total energy
spectra shown in Figure~\ref{fig:hypsp} are used to determine  $\alpha$.
Identifying, as usual, the eddy energy with the band-integrated
Fourier spectrum $\delta z_{\lambda}^2 \sim k_{\perp}\, E_{k_{\perp}}$,
where $k_{\perp} \sim \ell_{\perp}/\lambda$, from eq.~(\ref{eq:sbe1}) we obtain
\begin{equation}
E_{k_{\perp}} \propto k_{\perp}^{-\frac{3\alpha+2}{\alpha+2}},
\end{equation}
where for $\alpha = 1$ the $-5/3$ slope for the ``anisotropic Kolmogorov''
spectrum is recovered, and for $\alpha = 2$ the $-2$ slope for the
anisotropic IK case. At higher values of $\alpha$ correspond 
steeper spectral slopes up to the asymptotic value of $-3$.

\begin{figure}
     \includegraphics[width=0.47\textwidth]{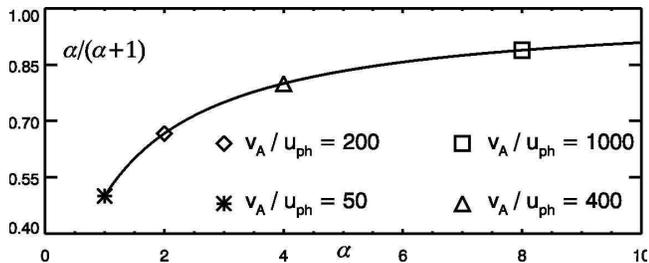}
          \caption{The solid line shows the exponent $\alpha/(\alpha+1)$ as a function
          of $\alpha$. Symbols  show values of $\alpha$ corresponding
          to different values of  $v_{\mathcal A}/u_{ph}$,
          at fixed $L/\ell_{\perp}=10$. \label{fig:spzl}}
\end{figure}

In Figure~\ref{fig:spzl} we plot the values of $\alpha$ determined in
this way,  together 
with the resulting power dependence $\alpha/(\alpha+1)$ of the 
amplitude~(\ref{eq:amp}) and of the energy flux~(\ref{eq:chs})  
on the parameter $f$.
The other power dependences are easily obtained from this last one,
e.g.\  for the energy flux~(\ref{eq:chs}) the power of the axial 
Alfv\'en speed $v_{\mathcal A}$ is given by $1+ \alpha/(\alpha+1)$,
so that in terms of the magnetic field $B_0$ it scales as $B_0^{3/2}$ for weak 
fields and/or long loops, to $B_0^2$ for strong fields 
and short loops.

Dividing eq.~(\ref{eq:chs}) by the surface $\ell_{\perp}^2$ we obtain the energy flux 
per unit area $F=\epsilon^{\ast}/\ell_{\perp}^2$.
Taking for example a coronal loop $40,000\ \textrm{km}$ long, 
with a  number density of 
$10^{10}\ \textrm{cm}^{-3}$, $v_{\mathcal A} = 2,000\ \textrm{km}\, \textrm{s}^{-1}$ and  
$u_{ph} = 1\ \textrm{km}\, \textrm{s}^{-1}$, 
(for these parameters we can estimate a value of $\alpha/(\alpha+1) \sim 0.95$),
which models an active region loop, we obtain 
$F \sim 5 \cdot {10}^6\ \textrm{erg}\, \textrm{cm}^{-2}\, \textrm{s}^{-1}$ 
and a Parker angle~(\ref{pan}) 
$<\Theta_P> \sim 4^{\circ}$. On the other hand, for
a coronal loop typical of a quiet Sun region, with a length of
 $100,000\ \textrm{km}$, a  number density of 
$10^{10}\ cm^{-3}$, $v_{\mathcal A} = 500\ \textrm{km}\, \textrm{s}^{-1}$ and  
$u_{ph} = 1\ \textrm{km}\, \textrm{s}^{-1}$, 
(for these parameters we can estimate a value of $\alpha/(\alpha+1) \sim 0.7$) 
we obtain $F \sim 7\cdot {10}^4\ \textrm{erg}\, \textrm{cm}^{-2}\, \textrm{s}^{-1}$ and 
$<\Theta_P> \sim 0.9^{\circ}$. 

In summary, with this paper we have shown how coronal heating rates in the 
Parker scenario scale with coronal loop and photospheric driving parameters,
demonstrating that field line tangling can supply the coronal heating energy 
requirement. We also predict that there is no universal scaling with axial 
magnetic field intensity, a feature which can be tested 
by observing weak field regions on the Sun, or the atmospheres 
of other stars with differing levels of magnetic activity.

\acknowledgements
M.V. thanks W.H.~Matthaeus for useful 
discussions. R.B.D. is supported by NASA SPTP.  \\

\end{document}